\documentclass[%
 reprint,
 amsmath,amssymb,
 aps,
]{revtex4-1}

\usepackage{graphicx}
\usepackage{dcolumn}
\usepackage{bm}
\usepackage{float}
\usepackage{xcolor}
\usepackage{etoolbox}
\apptocmd{\thebibliography}{\raggedright}{}{}

\begin{document}

\preprint{APS/123-QED}

\title {Pairing superbunching with compounded non-linearity in a resonant transition}

\author{G. Mouloudakis$^{1,2}$}
 \email{gmouloudakis@physics.uoc.gr}

\author{P. Lambropoulos$^{1,2}$}%

\affiliation{${^1}$Department of Physics, University of Crete, P.O. Box 2208, GR-71003 Heraklion, Crete, Greece
\\
${^2}$Institute of Electronic Structure and Laser, FORTH, P.O.Box 1527, GR-71110 Heraklion, Greece}

\date{\today}

\begin{abstract}
Through a quantitative analysis of an atomic transition driven strongly by quantized electromagnetic fields of various quantum states, we explore the role of quantum fluctuations on the behavior of the system. The emphasis is on fields with super-Poissonian statistics manifested in photon bunching, with the case of squeezed vacuum radiation serving as a prototype of superbuching. When combined with  non-linearly coupling of the resonant states,  bunching and superbunching lead to counter-intuitive behavior. The connection to recent progress in squeezed vacuum sources and the opportunity for experimental investigation, as well as challenging open theoretical problems are also outlined.  
\end{abstract}

\maketitle
The quantum stochastic properties of electromagnetic (EM) radiation, embodied in its correlation functions may arguably be viewed as the "trade mark" of quantum optics. As pointed out by Glauber \cite{ref1}  long ago, the set of correlation functions reflects the physical processes  that have generated the radiation.  Although Glauber's insightful remark referred to photons, spectacular developments since that time have proven its relevance to any bosons; be it in BEC, mesons, etc. The value of the intensity correlation functions of an atom beam from a Bose-Einstein condensate \cite{ref2,ref3} provide reassurance that the beam does indeed possess some of the minimal properties of a coherent state. At a much high higher energy of a few hundred GeV, in heavy ion nuclear collisions, two-particle correlation functions of $\pi$ or K mesons \cite{ref4,ref5}, provide a window into the degree of thermalization during the collision. Analogous, but not exactly the same, several particle (up to 4) angular correlations, in TeV gluon-quark matter \cite{ref6} are examined as a diagnostic tool of the state of the source.  

Apart from the "diagnostic" role of correlation functions, there is an equally fundamental manifestation of their properties, namely the dramatic and counterintuitive influence on non-linear radiation matter interaction.  Whereas, the outcome of a single-photon transition, driven by a weak field, depends only on the intensity (first order correlation function), even the simplest non-linear processes depend on higher order correlation functions. Two or more-photon absorption and harmonic generation \cite{ref7,ref8,ref9,ref10,ref11,ref12,ref13} represent the simplest examples, in which the yield depends on the intensity correlation function corresponding to the order of the respective non-linearity.  In a quite different context, the dynamics of even a two-level system (TLS) driven strongly by a single radiation mode, through a single-photon coupling, depends on correlation functions of essentially all orders \cite{ref14,ref15}. The observation of more general non-linear processes driven by stochastic fields, however, does in general require sufficiently high intensity, in combination with the ability to tailor its stochastic properties.  This dual requirement has been a hindrance in the experimental investigation of  theoretical predictions \cite{ref14,ref15} that have remained dormant over a period of essentially half a century. Owing to impressive advances in the engineering of states of radiation \cite{ref16,ref17}, however, that is no longer the case.  Most remarkably, the recent groundbreaking observation of up to fourth harmonic generated by squeezed light \cite{ref17} has broken that experimental barrier. At the same time, much recent interest has arisen in applications exploiting the role of photon statistics in non-linear radiation-matter interactions \cite{ref18,ref19,ref20,ref21,ref22}.

 A benchmark problem encapsulating the interplay of non-linearity and photon correlations is the dynamics of a transition driven resonantly by fields of different quantum stochastic processes, such as coherent state, chaotic (thermal) or squeezed. The dynamics can be probed either by the observation of the spectrum of spontaneous emission from the upper state or through a pump probe arrangement, as in double resonance (DR), where a second  weak-field transition to a third level serves as a probe. The case of a single-photon transition between two discreet levels, driven strongly by a coherent as well as a chaotic field,  has been studied in great detail, in the context of both resonance fluorescence and DR \cite{ref14,ref15}. Resonance fluorescence into a squeezed vacuum (SV) reservoir has been studied \cite{ref23,ref24,ref25,ref26,ref27,ref28,ref29,ref30,ref31} as has the case of the absorption spectrum in a squeezed state of non-zero bandwidth \cite{ref32,ref33}.  In addition to intensity fluctuations, a discreet-discreet transition is also sensitive to the bandwidth of the radiation. The bandwidth is of course part of the stochastic properties of the field, but not a uniquely quantum feature, as the intensity fluctuations are. For coherent and chaotic fields (CF), which are amenable to modelling in terms of classical fields, with Gaussian stochastic processes, this has been treated in considerable detail \cite{ref14,ref15,ref34}. The theory of a single-photon transition, driven strongly by coherent or CF, including arbitrary bandwidth, is well understood for both fluorescence and DR; although experimentally much remains to be explored, especially for driving by CF. To the best of our knowledge, no experimental data exist for that case, for which the theory predicts \cite{ref14,ref15,ref34} a rich variety of counter-intuitive effects due to intensity fluctuations mirrored in the concomitant fluctuations of the Rabi frequency. Given that photon bunching, inherent in chaotic radiation, is responsible for those effects, would those effects be more pronounced in the presence of superbunching?  In the light of the recent achievement by Spasibko et al. \cite{ref17}, theoretical predictions to those questions appear to be within experimental reach, which has served as the principal motivation for this paper.

The most conspicuous effect of intensity fluctuations on a strongly driven TLS is the distortion of the profile of AC Stark splitting. When probed in DR as a function of the probe frequency, in the absence of intensity fluctuations, the profile consists of two peaks separated by the Rabi frequency. But the Rabi frequency itself, is affected  by the intensity fluctuations, in a manner that depends on the order of the transition coupling the two levels and of course the stochastic properties of the field. One of the counter-intuitive predictions, that has never been tested  experimentally, is the disappearance of the splitting, when the TLS is driven by a transition of order higher than one, as in a two-photon transition.  In contrast to the single-photon transition, where the non-linearity sets in upon strong driving,  two or more photon strong driving involves a non-linearity imposed on an already non-linear process. It is this "escalation", so to speak, of non-linearity in combination with the stochastic character of the driving field that leads to counter-intuitive behavior. In order to single-out the fundamentally quantum aspect of intensity fluctuations, we have chosen to consider a source of zero bandwidth. To forestall misinterpretation of that condition, zero bandwidth here simply implies a source bandwidth sufficiently smaller than the width of the excited state; a situation which is well within experimental accessibility.  

Since squeezed light in not amenable to simulation in terms of classical stochastic processes, a fully quantum treatment involving averaging over the photon number distribution of the source is necessary. In order to keep the calculations computationally manageable, we have adopted appropriately scaled atomic parameters. Through parallel calculations with realistic parameters and laser intensities in earlier work \cite{ref35}, we have shown the equivalence of our scaled results with actual atomic systems and laser intensities. We have even shown the equivalence of our scaled parameters to recent experimental results \cite{ref17} in a completely different material system.

The atom is assumed to be initially in its ground state $\left|g\right\rangle$, with the radiation field in a two-mode number (Fock) state, with n photons of frequency $\omega_1$ in the first mode and m photons of frequency $\omega_2$ in the second one, denoted by $\left|n\right\rangle_1$ $\left|m\right\rangle_2$. Absorption of N photons of the first mode connects $\left|g\right\rangle$ to an excited state $\left| a \right\rangle$, which is in turn coupled to another excited state $\left| b \right\rangle$ via the  emission of a photon in the second radiation mode. The upper state $\left| a \right\rangle$ of the driven system is here assumed to be  energetically above state $\left| b \right\rangle$. The relevant states of the compound system (Atom + Radiation), expressed as tensor products of atomic and radiation field states are: $\left| I \right\rangle  = \left| g \right\rangle \left| n \right\rangle_1 \left| m \right\rangle_2$, $\left| A \right\rangle  = \left| a \right\rangle \left| n-N \right\rangle_1 \left| m \right\rangle_2$ and $\left| B \right\rangle  = \left| b \right\rangle \left| n-N \right\rangle_1 \left| m+1 \right\rangle_2$, with respective energies $\hbar{\omega _I} = \hbar{\omega _g} + n\hbar\omega_1 + m\hbar\omega_2 $, $\hbar{\omega _A} = \hbar{\omega _a} + (n-N)\hbar\omega_1 + m\hbar\omega_2 $ and $\hbar{\omega _B} = \hbar{\omega _b} + (n-N)\hbar\omega_1 + (m+1)\hbar\omega_2 $. The detunings from resonance for the two transitions are defined as $\Delta_1  \equiv N\omega_1  - {\omega _{ag}} \equiv N\omega_1  - ({\omega _a} - {\omega _g})$ and $\Delta_2  \equiv \omega_2  - {\omega _{ab}} \equiv \omega_2  - ({\omega _a} - {\omega _b})$. To account for the spontaneous decay of $\left|a\right\rangle$ to $\left|b\right\rangle$, as well as the possible ionization of $\left|a\right\rangle$ via the absorption of one additional photon, we  introduce the substitution ${\omega _A} \to {{\tilde \omega }_A} = {\omega _A} - \frac{i}{2}({\gamma _a}+{\Gamma _{ion}})$, with $\gamma_a$ denoting the spontaneous decay rate and $\Gamma_{ion}$ the ionization rate. The introduction of the spontaneous decay $\gamma_a$, without accounting for the repopulation of energetically lower levels is an approximation. However, since we focus on the regime  where the first transition is strong, with the induced Rabi frequency much larger than the spontaneous decay ${\gamma _a}$, the relative error due to this approximation is practically negligible.

The quantity needed for the calculation of the effect of photon statistics on the dynamics of the strongly driven levels is the population ${P_{b}}(n,t)$ of the probe final state $\left|b\right\rangle$ as a function of the detuning $\Delta_2$ of the probe frequency, averaged over the photon probability distribution for the strongly driving field state. For the calculation of ${P_{b}}(n,t)$ we employ a formulation of the time development of the system in terms of the resolvent operator $G(z-H)^{-1}$ with the Hamiltonian $H = H_{0}+H_{R}+V$ consisting of the sum of, respectively the atomic, radiation and interaction parts. Solving the equations of motion of the relevant matrix elements of the resolvent operator, as described in the supplementary material, we obtain ${P_{b}}(n,t)$ for $t>0$, which  is a function of the time as well as the photon number of the incident radiation field modes. The dependence on the photon numbers $n$ and $m$ comes from, both, the compound system energies $\hbar\omega_I$, $\hbar\tilde\omega_A$ and $\hbar\omega_B$, as well as the interaction Hamiltonian matrix elements $V_{AI}$ and $V_{BA}$. These matrix elements are related to the Rabi frequencies $\Omega_1^{(N)}$ and $\Omega_2$ of the $\left|g\right\rangle \leftrightarrow \left|a\right\rangle$ and $\left|a\right\rangle\leftrightarrow\left|b\right\rangle$ transitions, respectively, via the relations $\Omega_1^{(N)}=2V_{AI}$ and $\Omega_2=2V_{BA}$. Here, $\Omega_1^{(N)}$ represents an N-photon effective Rabi frequency and is therefore proportional to $\sqrt{n(n-1)...(n-N)} \approx n^N$ for large photon numbers, whereas $\Omega_2$ is simply proportional to $\sqrt{m}$, since it represents a single-photon transition. The ionization rate $\Gamma_{ion}$ introduced in the expression of $\tilde\omega_A$, depends linearly on n.

Our main objective is the behavior of  ${P_{b}}(n,t)$ averaged over the photon probability distribution of the driving source, as a function of $\Delta_{2}$,  with emphasis  on sources  prepared in states with super-Poissonian statistics. Specifically, we  consider the chaotic and the (bright) squeezed vacuum states, the results for which we compare to the case of the  field prepared in a coherent state. In spirit, this scheme is similar to the one employed in the treatment of the Jaynes-Cummings model found in many quantum optics textbooks \cite{ref36}. It is valid in our context of a source of zero bandwidth, in the sense defined in the previous section.  The averaging of $P_{b}(n,t)$ over the photon number distributions corresponding to the coherent, the chaotic and the SV states is formally given by the equations, 

\begin{equation}
P_{b}^{coh}(t) = \sum\limits_{n = 1}^\infty  {e^{ - \bar n}}\frac{{{{\bar n}^n}}}{{n!}}P_{b}(n,t)
\label{SumCoh}
\end{equation}
\begin{equation}
P_{b}^{chao}(t) = \sum\limits_{n = 1}^\infty  \frac{{{{\bar n}^n}}}{{{{(1 + \bar n)}^{n + 1}}}}P_{b}(n,t)
\label{SumChao}
\end{equation}
\begin{equation}
P_{b}^{SV}(t) = \sum\limits_{n = 1}^\infty  \frac{1}{{\sqrt {1 + \bar n} }}\frac{{(2n)!}}{{{{(n!)}^2}{2^{2n}}}}{\left( {\frac{{\bar n}}{{1 + \bar n}}} \right)^n}P_{b}(2n,t)
\label{SumSqVac}
\end{equation}
where ${\bar n}$ is the mean photon number. Note that the squeezed vacuum photon number distribution contains only even number of photons and therefore the increment of $P_{b}(n,t)$ in Eqn. (3) is 2$n$.

In order for the Stark splitting to develop in realistic transitions, we need to consider large mean photon numbers ($\bar n \approx 10^{6}$ or even much larger, as the order N of the coupling is increased). This entails a serious numerical task since the averages over the photon number distributions extend far beyond $\bar n$ and therefore, performing the summation through increments of n by 1 is computationally futile. One efficient way to overcome this obstacle, used also  in previous works of one of us (PL) \cite{ref37,ref38}, is to scale the problem by multiplying the photon number n by a dimensionless scaling factor in the expression of $P_{b}(n,t)$.  The scaling renders the averaging  numerically feasible, so that we can calculate the population profiles for values of $\bar n$, which correspond realistically to mean photon numbers around $\bar n$ times the scaling factor.  Owing to the non-linear dependence of $P_{b}(n,t)$ on n, the connection between realistic and scaled intensities is not straightforward. A discrepancy of a few percent \cite{ref38} that may be involved  is far from significant for the effects discussed in this paper.

The scaled parameters, such as Rabi frequencies, decay constants and photon numbers, employed in our calculations have been chosen so as to correspond to numbers compatible with realistic atomic systems and relevant laser intensities, guided by findings in an earlier publication \cite{ref35}. The value of the probe transition, which must be weak, can always be adjusted through the intensity of the probe, which means that the value of the relevant matrix element is of no consequence to our results.

The expressions that connect $\bar\Omega_1^{(N)}$ and $\bar\Gamma_{ion}$ to the intensity can easily be found using the relevant atomic parameters found in the literature \cite{ref39,ref40,ref41,ref42,ref43} or by calculating them. We have done both. We do in addition need the connection between laser intensity and the mean photon number employed in the calculation. For that purpose, we use the relation \cite{ref44} $\bar n=\frac{8 {\pi}^3 c^2}{{\omega}^2}\frac{F}{\gamma_L}$, where $\gamma_L$ is the bandwidth of the source, $\omega$ its frequency, and $F$ is the photon flux, connected to the intensity via the relation $F=I/N\hbar\omega$, where N is the order of the process. This equation involves an approximation in that it does not account for the detailed line shape of the source. It is, however, entirely sufficient for our  purpose in this paper, as it provides a relation between $\bar n$ and $I$ more or less within a factor of 2.

For the averaging over the photon distributions to be valid, the bandwidth of the source should be much smaller than the spontaneous decay width of state $\left|a\right\rangle$. To satisfy that condition, for both transitions, we choose a source bandwidth  100 times smaller than $\gamma_a$.
As mentioned above, in order to avoid using arbitrary parameters, for the single-photon strongly coupled system, our parameters correspond approximately to the levels $6s$ and $7p$ of Cesium, with the transition 7p $\rightarrow$ 7s serving as the probe. For the two-photon strongly coupled system, our parameters correspond to the two-photon transition 2s$\rightarrow$4s in Lithium, with the transition 4s$\rightarrow$3p as the probe. The ionization channels form the upper state, although not essential to the problem, have been included for the sake of completeness. Their presence or absence would depend on the atomic system chosen for experimental investigation of the effects discussed in this paper. The effective dipole matrix element of the two-photon transition was calculated  through the relation \cite{ref42} $\mu_{ga}^{(2)}=\sum_{l}\frac{\mu_{gl}\mu_{la}}{\omega_{la}+\omega_1}$.

\begin{figure}[H]
	\centering
	\includegraphics[width=8.5cm]{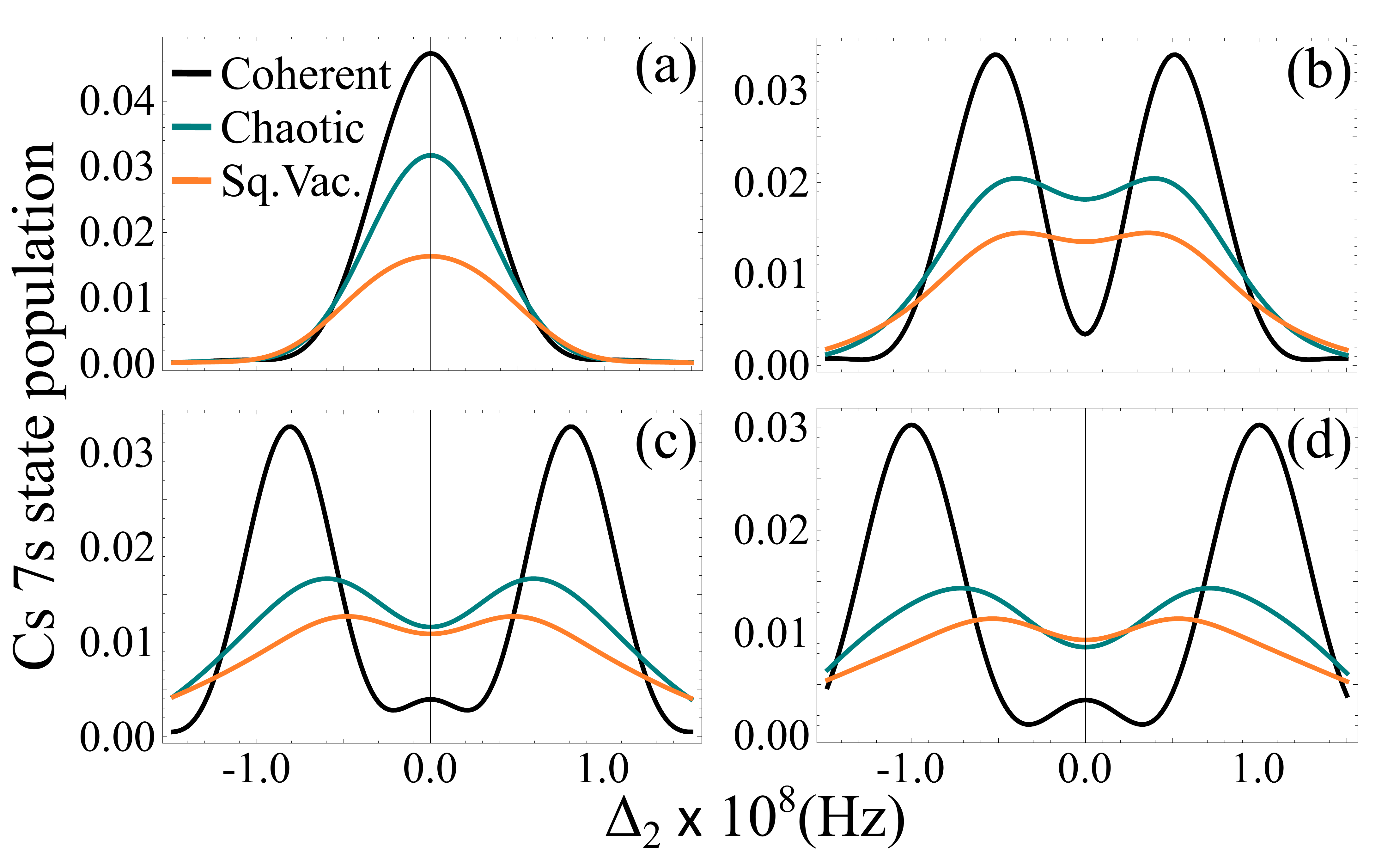}
		\caption[Fig.1]{Single-photon Stark splitting in Cesium. The calculation is carried out for various types of radiation fields: coherent (black line), chaotic (teal line) and squeezed vacuum (orange line).  The values of the relevant parameters used are: detuning from the first resonance $\Delta_1=0$, interaction time $T=10^{-7}$ sec, laser bandwidth $\gamma_{L}=0.82\times10^{5}$ Hz, scaling factor $=4\times10^6$, corresponding to the intensity: (a) $I=0.086$ W/cm$^{2}$, (b) $I=0.43$ W/cm$^{2}$, (c) $I=0.86$ W/cm$^{2}$, (d) $I=1.29$ W/cm$^{2}$.}
\end{figure}

In Figure 1 we present the population of the probe state as a function of $\Delta_2$, for the case of the TLS driven on resonance ($\Delta_1 =0$) by single photon coupling. It has been calculated for initial radiation fields with different quantum stochastic properties (i.e. coherent-black lines, chaotic-teal lines and squeezed vacuum-orange lines). For the comparison to be meaningful the average photon number $\bar n$ is the same for all field states. In the regime of intensities  corresponding to Rabi frequencies comparable to, or only  a few times larger than the natural width of the upper state (Figure 1a), increasing the intensity results to the broadening of the profile (power broadening) \cite{ref45}, until it eventually splits into two, forming the well known double peak structure (Autler-Towns doublet) \cite{ref46}, evident for all three initial field states (Figure 1b). For the sake of calibration, we note that for the transition 6s$\rightarrow$7p in Cs and the time of interaction assumed for Figure 1, the splitting would begin becoming noticeable around $I\approx 0.17$ W/cm$^{2}$; which is reasonable for a single-photon transition. For a coherent field state, the splitting is equal to the mean Rabi frequency $\bar \Omega_1$ of the driven single-photon transition, while for a chaotic field (CF) state, with increasing intensity, it tends to the value of $\frac{\bar \Omega_1}{\sqrt{2}}$, in agreement with earlier work \cite{ref14,ref15,ref34}. For an initial (bright) squeezed vacuum (SV) field state, as shown in Figures 1b to 1d, the splitting turns out to be even smaller than the one for chaotic field, at the same average intensity. In addition, both the chaotic and the SV fields cause a smearing out of the profile, as compared to that for the  coherent field.  Physically, this can be attributed to strong amplitude fluctuations, entailing  fluctuations of the Rabi frequency and therefore partial smearing of the doublet structure \cite{ref47}. It turns out that this effect is more pronounced for the SV as compared to the CF, which may be viewed as a reflection of the superbunching of SV, a term implying intensity fluctuations stronger than those of the CF. Thus, in general, the behavior of a single-photon bound-bound transition driven strongly by a SV field is in overall agreement with expectations based on the combination of the results of previous studies for the CF \cite{ref14,ref15,ref34} and the superbunching properties of the squeezed radiation. The differences in the behavior depicted in Figure 1 from those earlier results  are of a  quantitative but not qualitative degree. 

The situation changes drastically when a strongly driven two-photon transition is examined, as illustrated with the results shown in Figure 2. For weak to moderate driving field strengths, again a gradual (power) broadening \cite{ref48} appears (Figure 2a), developing eventually into a doublet. Under stronger driving, however, at least two glaring surprises stand out. First, in contrast to the single-photon case, with increasing intensity, the splitting  present under driving by a coherent state, is totally absent under driving by  chaotic or squeezed vacuum fields, giving rise to a single peak (Figure 2b). Second, the width of that peak is significantly smaller than the average Rabi frequency. Moreover, inspection of Figure 2d reveals that the width of the peak for SV driving, if anything tends to be smaller than the one for CF driving; a rather unexpected feature since SV undergoes stronger fluctuations, in the sense that its intensity correlation functions are larger \cite{ref49,ref17,ref50}. This highly counter-intuitive effect persists for even higher intensities, corresponding to mean Rabi frequencies many times larger than than spontaneous decay of the excited state (Figures 2c and 2d). Its physical interpretation defies straightforward extrapolation from the physical picture for single to two-photon transition, pointing instead to an intricate interplay between the non-linearity of the transition itself and that induced by the strong driving; hence our term "compounded non-linearirty". Obviously it is not limited to two-photon driving, promising  more intriguing features for three-photon, which as demonstrated by the results of Spasibko et al. \cite{ref17} is well within reach of current experimental possibilities. 

In undertaking this work, our expectation was that driving an initially non-linear process by superbunched radiation, would result to a spectrum in DR smeared over the range of the mean Rabi frequency, due to the strong fluctuation of the instantaneous Rabi frequency induced by the fluctuations of a superbunched source; an intuition based on earlier work for driving by chaotic (bunched) radiation \cite{ref14,ref34}. Obviously the underlying physics is much more complex, whereas these issues are now within experimental reach, which has eluded the field for a long time, posing at the same time a number of challenging theoretical problems, such as for example the role of the bandwidth under strong driving of a non-linearly coupled resonance by a superbunched source.   

\begin{figure}[H]
	\centering
	\includegraphics[width=8.5cm]{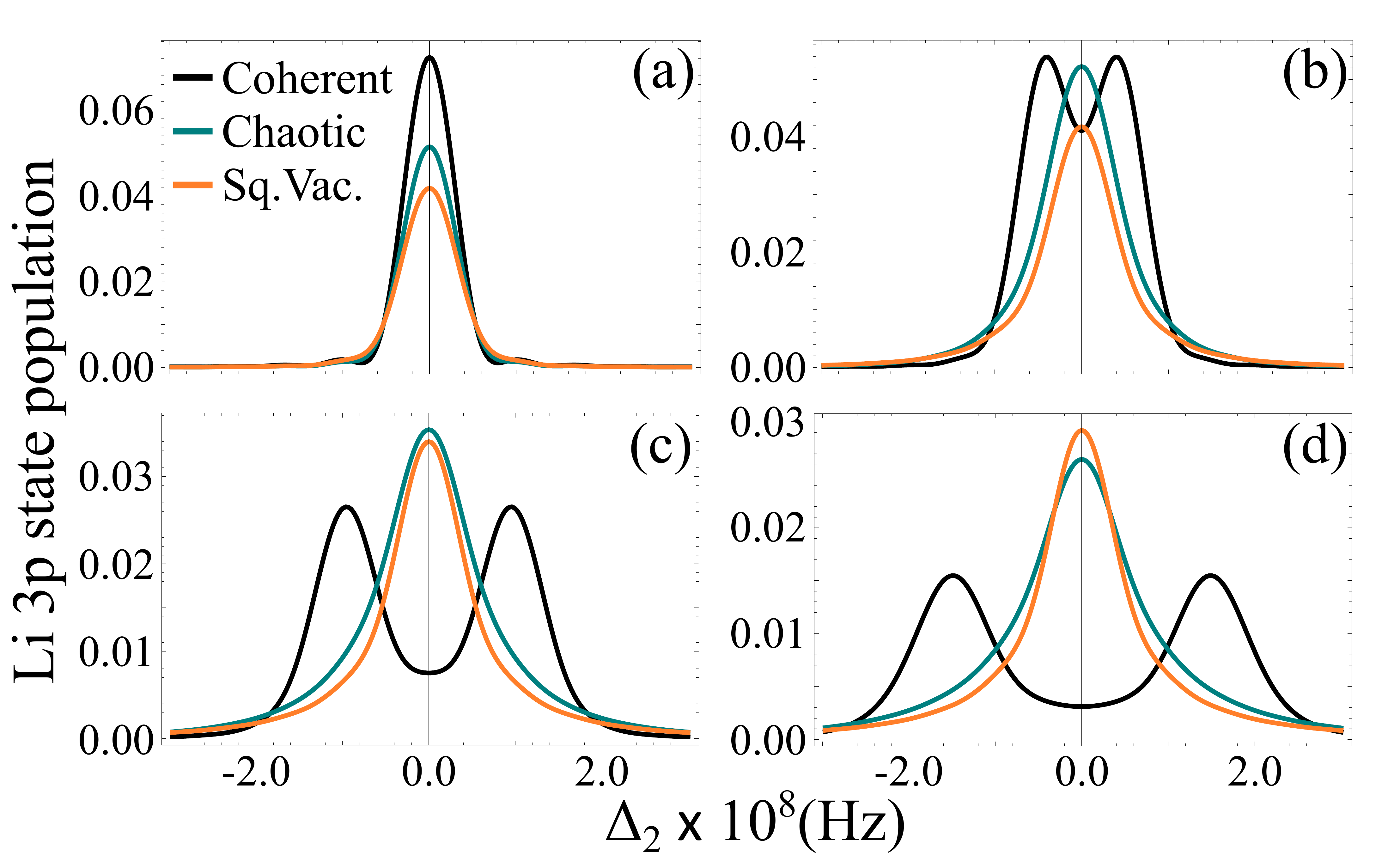}
	\caption[Fig.2]{Two-photon Stark splitting in Lithium. The calculation is carried for various types of radiation fields: coherent (black line), chaotic (teal line) and squeezed vacuum (orange line). The values of the relevant parameters used are: detuning from the first resonance $\Delta_1=0$, interaction time $T=10^{-7}$ sec, laser bandwidth $\gamma_{L}=1.78\times10^{5}$ Hz, scaling factor $=10^{13}$, corresponding to the intensity: (a) $I=0.49\times10^{6}$ W/cm$^{2}$, (b) $I=2.47\times10^{6}$ W/cm$^{2}$, (c) $I=4.95\times10^{6}$ W/cm$^{2}$, (d) $I=7.43\times10^{6}$ W/cm$^{2}$.}
\end{figure}

 GM thankfully acknowledges financial support by the  program  $\Pi \Delta$E$00710$ of the Institute of Electronic Structure and Laser, FORTH.

\newpage

\section* {Supplementary Material}

The theory is cast in terms of the resolvent operator, defined by $G(z)\equiv (z-H)^{-1}$ [44], where $H$ is the total Hamiltonian of the system, consisting of three parts, i.e. the atomic Hamiltonian $H_0$, the Hamiltonian of the radiation field $H_R$ and the interaction  $V$ between the two subsystems. As depicted in Figure 1, the atom is assumed to be in its ground state $\left| g \right\rangle$ and the radiation field  prepared in a two-mode number (Fock) state, denoted by $\left|n\right\rangle_1$ with n photons of frequency $\omega_1$ in the first mode and $\left|m\right\rangle_2$ with m photons of frequency $\omega_2$ in the second one. Absorption of N photons from the first mode leads the atom to an excited state denoted by $\left| a \right\rangle$. This upper state is coupled to another energetically lower excited state $\left| b \right\rangle$, via the emission of a photon into the second radiation mode. The relevant states of the compound system (Atom + Radiation) expressed in terms of tensor products of atomic and radiation field states are: $\left| I \right\rangle  = \left| g \right\rangle \left| n \right\rangle_1 \left| m \right\rangle_2$, $\left| A \right\rangle  = \left| a \right\rangle \left| n-N \right\rangle_1 \left| m \right\rangle_2$ and $\left| B \right\rangle  = \left| b \right\rangle \left| n-N \right\rangle_1 \left| m+1 \right\rangle_2$, with energies $\hbar{\omega _I} = \hbar{\omega _g} + n\hbar\omega_1 + m\hbar\omega_2 $, $\hbar{\omega _A} = \hbar{\omega _a} + (n-N)\hbar\omega_1 + m\hbar\omega_2 $ and $\hbar{\omega _B} = \hbar{\omega _b} + (n-N)\hbar\omega_1 + (m+1)\hbar\omega_2 $, respectively. We also define the detunings from resonance of the two transitions as $\Delta_1  \equiv N\omega_1  - {\omega _{ag}} \equiv N\omega_1  - ({\omega _a} - {\omega _g})$ and $\Delta_2  \equiv \omega_2  - {\omega _{ab}} \equiv \omega_2  - ({\omega _a} - {\omega _b})$ and make the substitution ${\omega _A} \to {{\tilde \omega }_A} = {\omega _A} - \frac{i}{2}({\gamma _a}+{\Gamma _{ion}})$ to account for the spontaneous decay $\gamma_a$, as well as the ionization of state $\left| a \right\rangle$ via the absorption of an additional photon. Note that this method of introducing the spontaneous decay is approximate in the sense that it does not take into account the repopulation of energetically lower levels caused by it. However, since we  focus in the regime where the first transition is strong and therefore, the induced Rabi frequency of the first transition is much greater than the spontaneous decay ${\gamma _a}$, the relative error caused by this method is practically negligible.

\begin{figure}[H]
	\centering
	\includegraphics[width=7cm]{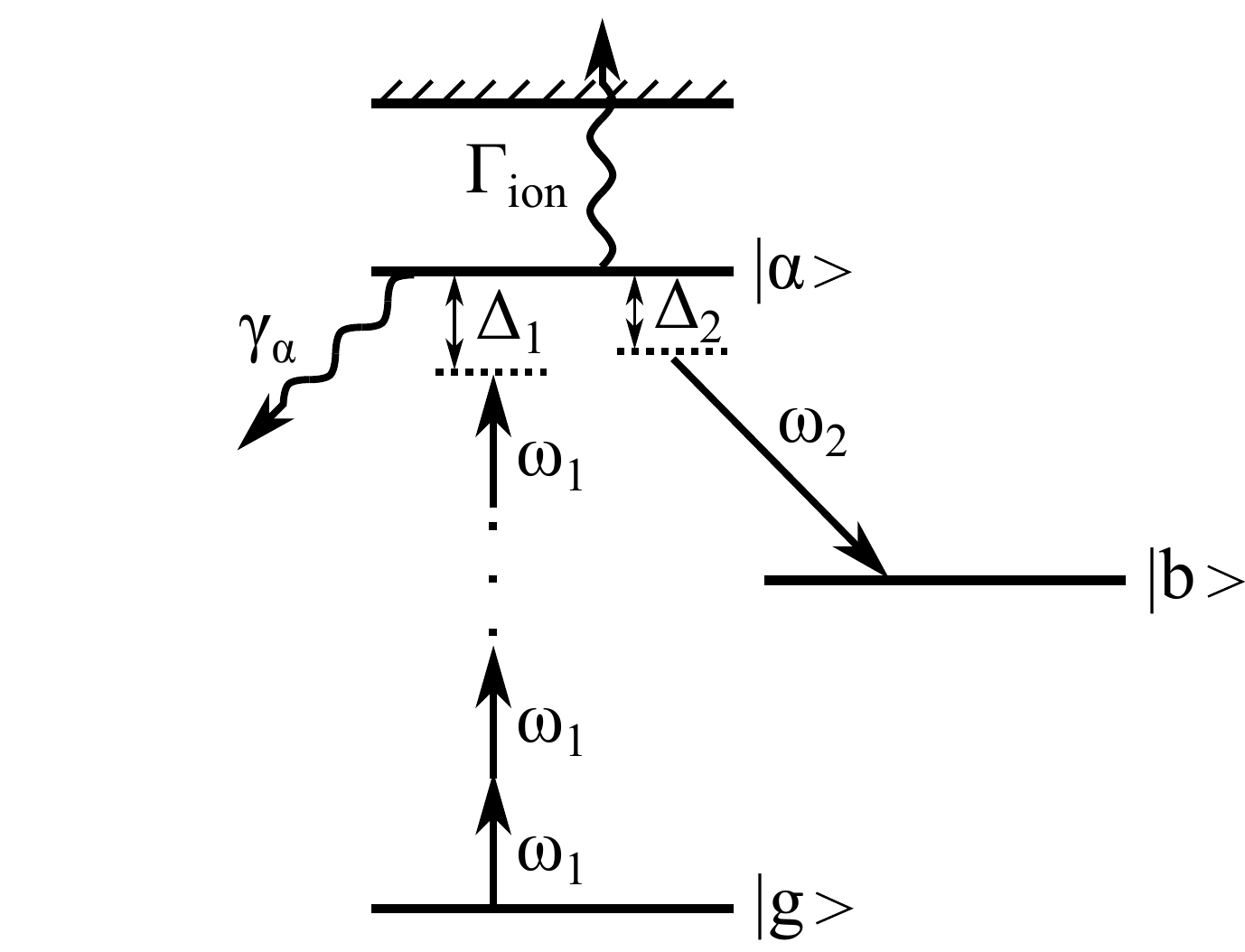}
		\caption[Fig.1]{Schematic presentation of the system under consideration}
\end{figure}

The equations of motion of the resolvent operator matrix elements in the compound system basis are:
\begin{equation}
(z - {\hbar\omega _I}){G_{II}} = 1 + {V_{IA}}{G_{AI}}
\end{equation}
\begin{equation}
(z - {\hbar\tilde\omega _{\rm A}}){G_{AI}} = {V_{AI}}{G_{II}} + {V_{AB}}{G_{BI}} 
\end{equation}
\begin{equation}
(z - {\hbar\omega _B}){G_{BI}} = {V_{BA}}{G_{AI}}
\end{equation}

Solving the above system of equations for ${G_{BI}}$, one obtains:
\begin{widetext}
\begin{equation}
G_{BI}=\frac{V_{BA}V_{AI}}{(z-\hbar\omega_I)(z-\hbar\tilde\omega_A)(z-\hbar\omega_B)-(z-\hbar\omega_I){\left|V_{BA}\right|}^2-(z-\hbar\omega_B){\left|V_{AI}\right|}^2}
\label{GBI}
\end{equation}
\end{widetext}

The time evolution operator matrix elements ${U_{ij}}(t)$ are related to the respective resolvent operator's matrix elements via the equation [44]
\begin{equation}
{U_{ij}}(t) =  - \frac{1}{{2\pi i}}\int_{ - \infty }^{ + \infty } {{e^{ - ixt}}{G_{ij}}({x^ + })dx} 
\end{equation}
where ${x^ + } = x + i\eta $, with $\eta  \to {0^ + }$. This Laplace inversion  integral is obtained in terms of the roots of the third order polynomial appearing in the denominator of Eq. (\ref{GBI}). Denoting these three roots by $z_1$, $z_2$ and $z_3$, the resulting expression is
\begin{equation}
\begin{aligned}
{U_{BI}}(t) = & {V_{BA}}{V_{AI}}  \left[ \frac{{\exp ( - iz_{1}t)}}{{(z_1 - {z_2})(z_1-z_3)}} +  \right. \\
& \left. \frac{{\exp ( - i{z_2}t)}}{{({z_2} - {z_1})({z_2} - {z_3})}} + \frac{{\exp ( - i{z_3}t)}}{{({z_3} - z_1)({z_3} - {z_2})}} \right]
\end{aligned}
\label{UBI}
\end{equation}
The population of state $\left|b\right\rangle$  at times $t>0$, is given by
\begin{equation}
{P_{b}}(t) = {\left| {{U_{BI}}(t)} \right|^2}
\end{equation}
Note that the population is a function of the time as well as the photon number of the  radiation field modes. The dependence on the photon numbers $n$ and $m$, comes both from the compound system energies $\hbar\omega_I$, $\hbar\tilde\omega_A$ and $\hbar\omega_B$, as well as the interaction Hamiltonian matrix elements $V_{AI}$ and $V_{BA}$. These matrix elements are related to the Rabi frequencies $\Omega_1^{(N)}$ and $\Omega_2$ of the $\left|g\right\rangle \leftrightarrow \left|a\right\rangle$ and $\left|a\right\rangle\leftrightarrow\left|b\right\rangle$ transitions, respectively, via the relations $\Omega_1^{(N)}=2V_{AI}$ and $\Omega_2=2V_{BA}$. $\Omega_1^{(N)}$ represents an N-photon effective Rabi frequency and is therefore proportional to $\sqrt{n(n-1)...(n-N)} \approx n^N$ for large photon numbers, whereas $\Omega_2$ is just proportional to $\sqrt{m}$, since it represents a single-photon transition. The ionization rate $\Gamma_{ion}$ introduced in the expression of $\tilde\omega_A$, also depends linearly on n.  In what follows we will be interested in the dependence of $P_{b}(t)$ solely on n and t, and therefore we adopt the notation $P_{b}(n,t)$ for the population of state $\left|b\right\rangle$.

If the first transition is strong, i.e. if $\Omega_1^{(N)}\gg$ $\gamma_a$, $\Gamma_{ion}$, $\Delta_1$, the interaction $V$ between the uncoupled ``atom+field" states $\left|I\right\rangle$ and $\left|A\right\rangle$, lifts  the degeneracy and causes them to split into doublets, an effect widely known as Stark splitting. In case of exact resonance ($\Delta_1=0$), which is the case studied in the paper, the ``new" eigenstates of the system (dressed atom states), are energetically separated by $\hbar\Omega_1$. This splitting is obtained  by calculating the population of state $\left|b\right\rangle$ as a function of $\Delta_2$, which  assumed to be weakly coupled to $\left|a\right\rangle$ and acts as a probe.

According to a method that can be found in many quantum optics textbooks [36], solving a problem with the radiation field initially prepared in a Fock state and obtaining a desired quantity as a function of the photon number, one can average over the photon number distributions of field states other than Fock and find the resulting quantity in case where the initial field is prepared in such radiation states. This method is sorely helpful to study the effects of different radiation field states on atomic transitions, but one should always be aware that it is only valid in the ``zero bandwidth" limit, i.e. when the field bandwidth is much smaller than the natural decay of the atomic states involved in the calculation. Adopting this method, we average $P_{b}(n,t)$ over the photon number distributions corresponding to a coherent, a chaotic and a squeezed vacuum (SV) initial radiation field state:

\begin{equation}
P_{b}^{coh}(t) = \sum\limits_{n = 1}^\infty  {e^{ - \bar n}}\frac{{{{\bar n}^n}}}{{n!}}P_{b}(n,t)
\label{SumCoh}
\end{equation}
\begin{equation}
P_{b}^{chao}(t) = \sum\limits_{n = 1}^\infty  \frac{{{{\bar n}^n}}}{{{{(1 + \bar n)}^{n + 1}}}}P_{b}(n,t)
\label{SumChao}
\end{equation}
\begin{equation}
P_{b}^{SV}(t) = \sum\limits_{n = 1}^\infty  \frac{1}{{\sqrt {1 + \bar n} }}\frac{{(2n)!}}{{{{(n!)}^2}{2^{2n}}}}{\left( {\frac{{\bar n}}{{1 + \bar n}}} \right)^n}P_{b}(2n,t)
\label{SumSqVac}
\end{equation}
Note that for an N-photon process all $P_{b}(n<N,t)$ terms are zero due to the presence of $V_{AI}$ in the numerator of Eqn. (\ref{UBI}) and therefore the first non-zero term in Eqns. (\ref{SumCoh}) and (\ref{SumChao}) is the $n=N$ term. Similarly, in the summation over the squeezed vacuum photon number distribution, the first non-zero term is the $n=N/2$ or $n=(N+1)/2$ term, depending on whether N is even or odd, respectively. The SV photon number distribution contains only even number of photons and therefore the increment of $P_{b}(n,t)$ in Eqn. (\ref{SumSqVac}) is 2$n$.

In order to maintain  a connection of our calculations with  transitions in actual physical systems, for our scaled parameters, we have chosen the atoms of Cesium and Lithium as systems of reference. In that context, we need to find the expressions connecting $\bar\Omega_1$ and $\bar\Gamma_{ion}$ (mean Rabi frequency and mean ionization rate) to $\bar n$, pertaining to the 6s $\rightarrow$ 7p and 2s $\rightarrow$ 4s transitions in Cs and Li, respectively, including the ionization channels. Note that we do not need to calculate the exact value of the matrix element connecting $\left|a\right\rangle$ to $\left|b\right\rangle$ (7p to 7s in Cs and 4s to 3p in Li), since $\left|b\right\rangle$ serves as a probe (weak coupling), and therefore we only care about the exact value of $\bar\Omega_{2}$ connecting the two excited resonances, and not the value of $m$. In other words, since the averaging does not include summation with respect to $m$, one may always choose the proper $m$, which means adjusting the probe laser intensity so as to obtain desired weak Rabi frequency $\bar\Omega_2$, irrespective of the value of the atomic matrix element. 

The relations that connect $\bar\Omega_1^{(N)}$ and $\bar\Gamma_{ion}$ to intensity can be found by calculating the corresponding atomic parameters [39-43]. Then, it is only the connection between the intensity and the mean photon number that is needed, in order to obtain them as function of the latter. An approximate connection between the two is given by [44] $\bar n=\frac{8 {\pi}^3 c^2}{{\omega}^2}\frac{F}{\gamma_L}$, where $\gamma_L$ is the bandwidth of the source, $\omega$ its frequency, and $F$ is the photon flux, connected to the intensity via the relation $F=I/N\hbar\omega$, with N being the order of the process. The approximation inherent in this equation is that it does not account for the exact line shape of the source.  It is, however, sufficient for our purposes of this paper, as it provides a relation between $\bar n$ and $I$ to within more or less  a factor of 2. As described in the main text, in order for the averaging over the photon distributions to be valid, the bandwidth of the source should be much smaller than the natural decay of state $\left|a\right\rangle$. For that reason, we have chosen for both transitions,  a bandwidth  100 times smaller than $\gamma_a$.  The resulting parameters used for the  6s $\rightarrow$ 7p (7s probe) transition in Cs are: $\omega_g=0$, $\omega_a=2.698$eV, $\omega_b=2.298$eV, $\Delta_1=0$, $\gamma_a=0.82\times10^7$Hz, $\bar\Omega_1=1.884\times10^4 \sqrt{\bar n}$ Hz, $\bar\Gamma_{ion}=1.181\times10^{-7} \bar n$ Hz, $\gamma_L=0.82\times10^5$ Hz, while for the 2s $\rightarrow$ 4s (3p probe) transition in Li are: $\omega_g=0$, $\omega_a=4.372$eV, $\omega_b=3.835$eV, $\gamma_a=1.78\times10^7$Hz, $\Delta_1=0$, $\bar\Omega_1=11.225\times10^{-7} \bar n$ Hz, $\bar\Gamma_{ion}=3.09\times10^{-7} \bar n$ Hz, $\gamma_L=1.78\times10^5$ Hz. In both transitions the Rabi frequency connecting the excited state to the probe state is chosen to be equal to the spontaneous decay of the former.  The effective two-photon dipole matrix element for the transition was calculated using the values of the single-photon dipole moments between allowed transitions in Li, via the relation [42] $\mu_{ga}^{(2)}=\sum_{l}\frac{\mu_{gl}\mu_{la}}{\omega_{la}+\omega_1}$. Using now the aforementioned relation between the intensity and the mean photon number we obtain the expression for $\bar\Omega_1$ as a function of $\bar n$.

A few comments should be made at this point. First of all, the inclusion of the ionization channels in both transitions stems from the fact that the zero detuning from resonance of the first transition ($\Delta_1=0$) implies that $\omega_1=2.698$ eV in Cs and $\omega_1=2.186$ eV in Li. Since the ionization thresholds lie at 3.894 eV and 5.392 eV in Cs and Li, respectively, the absorption of an additional photon of the first mode, while the atom is at the excited state $\left|a\right\rangle$, leads to ionization in both cases. 

Another comment is related to the fine structure splitting of the 7p state in Cs. Since we are working with a bandwidth much smaller than the fine structure splitting of the p line, which lies in the THz range, the spectral resolution of the source allows us to resolve such a splitting. In that case, a complete theoretical description should include both $J=1/2$ and $J=3/2$ states. However, since we tune exactly on resonance ($\Delta_1=0$) with one of the states of the doublet (take for example the $J=1/2$ resonance), the effect of the neighbouring state is negligible and it can be safely neglected. Similarly, the fine splitting of the 3p probe line in Li does not affect our results, since we consider detunings $\Delta_{2}$ in the vicinity of $10^{8}$ Hz, much lower than the fine structure  splitting of the line.

Finally, we should mention that the description of the 2s $\rightarrow$ 4s transition in Li, in terms of an effective two-photon Rabi frequency is, in general, not straightforward since the intermediate resonances between the two states may add additional complications to the system. This generalisation is only valid as long as the first photon that participates in the two-photon process is sufficiently off-resonance from the nearest intermediate state. In our case the first photon with energy $\omega_1=2.186$ eV is detuned by $\Delta_{2p}=0.338$ eV $\simeq 81$ THz from the 2p resonance in Li, which is sufficiently larger than the values of the scaled two-photon Rabi frequency considered (in the vicinity of $10^{8}$ Hz), ensuring the validity of our assumption.
\\

\textbf{Supplementary Material References}

36. See, for example, P. Lambropoulos and D. Petrosyan, \textit{Fundamentals of Quantum Optics and Quantum Information}, (Springer Verlag 2007)

39. H. Eicher, IEEE Journal of Quantum Electronics \textbf{11}, 4 (1975)

40. M. Aymar, E. Luc-Koenig and F. Combet Farnoux, J. Phys. B: Atom. Molec. Phys. \textbf{9}, 1279 (1976)

41. Shahid Hussain, M. Saleem, and M. A. Baig, Phys. Rev. A \textbf{74}, 052705 (2006)

42. R. J. Wolff and S. P. Davis, J. Opt. Soc. Am. \textbf{58}, 490-495 (1968)

43. R. W. Boyd, J. G. Dodd, J. Krasinski and C. R. Stroud, Opt. Lett. \textbf{5}, 117-119 (1980)

44. P. Lambropoulos, Adv. At. Mol. Phys. \textbf{12}, 87 (1976)

\end{document}